\documentclass[iop]{emulateapj}
\usepackage{amssymb,natbib,graphicx,epstopdf,color,lineno}

\newcommand{\Ha}{\ensuremath{\rm H\alpha} }
\newcommand{\LHa}{$L_{{\rm H}\alpha}$ }
\newcommand{\Hb}{\ensuremath{\rm H\beta} }
\newcommand{\HaHb}{\ensuremath{\rm H\alpha/\rm H\beta} }

\newcommand{\HII}{\ion{H}{2} }

\newcommand{\OIII}{[\ion{O}{3}] }
\newcommand{\OII}{[\ion{O}{2}] }

\newcommand{\OIIIb}{[\ion{O}{3}]$\lambda 5007$ }
\newcommand{\SII}{[\ion{S}{2}] }
\newcommand{\NII}{[\ion{N}{2}] }
\newcommand{\NIIa}{[\ion{N}{2}]$\lambda 6548$ }
\newcommand{\NIIb}{[\ion{N}{2}]$\lambda 6583$ }

\def\eg{{e.g.,~}}
\def\ie{{i.e.,~}}

\begin{document}


\title{DUST EXTINCTION FROM BALMER DECREMENTS OF STAR-FORMING GALAXIES AT $0.75\le z \le 1.5$ WITH HST/WFC3 SPECTROSCOPY FROM THE WISP SURVEY}

\author{{\sc A. Dom\'inguez}\altaffilmark{1}, {\sc B. Siana}\altaffilmark{1}, {\sc A.~L. Henry}\altaffilmark{2}, {\sc C. Scarlata}\altaffilmark{3}, {\sc A.~G. Bedregal}\altaffilmark{3}, {\sc M. Malkan}\altaffilmark{4}, {\sc H. Atek}\altaffilmark{5}, {\sc N.~R. Ross}\altaffilmark{4}, {\sc J.~W. Colbert}\altaffilmark{5}, {\sc H.~I. Teplitz}\altaffilmark{6}, {\sc M. Rafelski}\altaffilmark{6}, {\sc P. McCarthy}\altaffilmark{7}, {\sc A. Bunker}\altaffilmark{8}, {\sc N.~P. Hathi}\altaffilmark{7}, {\sc A. Dressler}\altaffilmark{7}, {\sc C.~L. Martin}\altaffilmark{2}, {\sc D. Masters}\altaffilmark{1,7}}

\slugcomment{Draft; \today}

\shorttitle{DUST EXTINCTION OF STAR-FORMING GALAXIES AT $0.75\le z \le 1.5$}
\shortauthors{DOM\'INGUEZ ET AL.}

\altaffiltext{1}{Department of Physics \& Astronomy, University of California Riverside, Riverside, CA 92521, USA}
\altaffiltext{2}{Department of Physics, University of California, Santa Barbara, CA 93106, USA}
\altaffiltext{3}{Minnesota Institute for Astrophysics, University of Minnesota, Minneapolis, MN 55455, USA}
\altaffiltext{4}{Department of Physics \& Astronomy, University of California Los Angeles, Los Angeles, CA 90095, USA}
\altaffiltext{5}{Spitzer Science Center, Caltech, Pasadena, CA 91125, USA}
\altaffiltext{6}{Infrared Processing and Analysis Center, Caltech, Pasadena, CA 91125, US}
\altaffiltext{7}{Observatories of the Carnegie Institution for Science, Pasadena, CA 91101, USA}
\altaffiltext{8}{Department of Physics, Denys Wilkinson Building, Keble Road, Oxford OX1 3RH}

\email{albertod@ucr.edu}

\begin{abstract}
Spectroscopic observations of \Ha and \Hb emission lines of 128 star-forming galaxies in the redshift range $0.75\le z \le 1.5$ are presented. These data were taken with slitless spectroscopy using the G102 and G141 grisms of the Wide-Field-Camera~3 (WFC3) on board the Hubble Space Telescope as part of the WFC3 Infrared Spectroscopic Parallel (WISP) survey. Interstellar dust extinction is measured from stacked spectra that cover the Balmer decrement (H$\alpha$/H$\beta$). We present dust extinction as a function of \Ha luminosity (down to $3\times 10^{41}$~erg~s$^{-1}$), galaxy stellar mass (reaching $4\times 10^{8}$~M$_{\odot}$), and rest-frame \Ha equivalent width. The faintest galaxies are two times fainter in \Ha luminosity than galaxies previously studied at $z\sim 1.5$. An evolution is observed where galaxies of the same \Ha luminosity have lower extinction at higher redshifts, whereas no evolution is found within our error bars with stellar mass. The lower \Ha luminosity galaxies in our sample are found to be consistent with no dust extinction. We find an anti-correlation of the [\ion{O}{3}]$\lambda 5007$/H$\alpha$ flux ratio as a function of luminosity where galaxies with $L_{H\alpha}<5\times 10^{41}$ erg~s$^{-1}$ are brighter in [\ion{O}{3}]$\lambda 5007$ than H$\alpha$.  This trend is evident even after extinction correction, suggesting that the increased [\ion{O}{3}]$\lambda 5007$/H$\alpha$ ratio in low luminosity galaxies is likely due to lower metallicity and/or higher ionization parameters.

\end{abstract}


\section{Introduction}
Star-forming galaxies are characterized by emission lines from gas heated by young stars in \HII regions. The luminosity of these \emph{nebular} emission lines such as \Ha or \OIII provides reliable information concerning galaxy star-formation rates (SFRs) and gas-phase metallicities. Yet the dust present in the interstellar medium strongly attenuates rest-frame ultraviolet and optical fluxes in a wavelength-dependent manner. Therefore, inferring physical properties (\eg SFRs) from emission-line luminosities requires accurate dust corrections.

In practice, the most reliable technique to estimate interstellar extinction is to measure the flux ratio of two nebular Balmer emission lines such as \HaHb (\ie the Balmer decrement). Since the value of the Balmer decrement is set by quantum physics, any deviation from this expected value may be attributed to dust extinction (for a fixed electron temperature). Other emission-line ratios, such as H$\beta$/H$\delta$ are useful as well, but are generally more difficult to measure because the higher order lines are weaker and more affected by stellar absorption.

The determination of dust extinction from the Balmer decrement has been shown to be a very successful technique in the local Universe since the first statistical work by \citet{kennicutt92}. These results have been improved upon by the large amount of optical spectra provided by the Sloan Digital Sky Survey (SDSS), which were analyzed in this context by \citet{brinchmann04}, \citet{moustakas06}, and \citet{garn10}. However, the simultaneous detection of H$\alpha$ and H$\beta$ in higher redshift galaxies, which are shifted to the near-infrared (IR), is a difficult task with current instrumentation on ground-based telescopes. Indeed, Balmer decrements in high-redshift galaxies have been spectroscopically measured only in a few individual cases (\eg \citealt{teplitz00,hainline09}).

Other recent, although less robust, estimates of the Balmer decrement are available at $z\sim 0.5$ for larger galaxy samples based on a combination of spectroscopy and photometry (\citealt{ly12}). There are other dust-extinction estimates at $z\sim 1.47$ from narrow-band photometry of the \Ha and \OII emission lines (\citealt{sobral12b}), but these measurements are affected by the unknown metallicities of the galaxies. At high redshifts ($z>2$) the ultraviolet (UV) slope is typically used to determine dust extinction (\eg \citealt{stanway05,hathi08,reddy10,finkelstein11,wilkins11,bouwens12,hathi12}), where access to the Balmer lines is difficult. The relationship between UV slope and dust reddening depends both on the dust properties and distribution of the dust with respect to the stars.  While the local correlation between UV slope and dust attenuation appears to hold for typical star-forming galaxies up to $z\sim 2$ (\eg \citealt{reddy10,reddy12}), the Balmer decrement offers a complementary method of deducing dust attenuation, and can illuminate any differences in the nebular versus stellar extinction in galaxies (\citealt{calzetti94,calzetti00}).

Previous works on dust extinction have focused on a relatively narrow dynamical range of \Ha luminosity (\eg \citealt{sobral12b}). Furthermore, these works have been generally sensitive to the brightest star-forming galaxies, which tend to have larger reddening than fainter galaxies. As a result, the common procedure to correct for dust extinction in high-redshift galaxies is to assume an extinction of one magnitude at the wavelength of H$\alpha$ for all galaxy luminosities. This assumption is based on the mean extinction of local galaxies (\citealt{kennicutt92}). However, this local average has a significant scatter, which may potentially affect results derived from the dust correction. An additional issue is that dust extinction as a function of bolometric luminosity is observed to evolve with redshift such that galaxies of the same luminosity are more extinguished at lower redshift, due mainly to the metal enrichment of the interstellar medium (\eg \citealt{reddy06,buat07,burgarella07,bouwens09,reddy10,buat11,bouwens12}).

The Wide Field Camera 3 (WFC3) of the Hubble Space Telescope (HST) has now made it possible to carry out a statistically significant analysis of the Balmer decrement in high-redshift galaxies. The grism spectroscopy mode of the WFC3 instrument offers several advantages over ground-based spectroscopy. First, grism spectroscopy is slitless, so there are no concerns of variable slit losses at H$\alpha$ and H$\beta$ due to varying seeing, alignment errors, and atmospheric refraction. Second, the WFC3 grisms provide a continuous wavelength coverage over $0.85\le \lambda \le 1.65$~$\mu$m, which is unaffected by hydroxyl (OH) airglow emission and the high opacity of the atmosphere at certain wavelengths (\ie line measurements are obtained regardless of whether they lie within atmospheric windows or between sky lines). Third, the flux calibration with WFC3 is more accurate than ground-based instruments, as it is unaffected by variable telluric absorption from Earth's atmosphere. Lastly, the WFC3 grisms obtain unbiased samples, whereas ground-based spectroscopy typically requires a preselection in order to assign slits. Furthermore, longslit followup at higher redshifts has mostly focused on brighter galaxies.

In this paper, interstellar dust-extinction of star-forming galaxies is statistically investigated in the redshift range $0.75\le z \le 1.5$ over two orders of magnitude in \Ha luminosities. Balmer decrements are measured using spectroscopy and analyzed down to an \Ha flux limit of approximately $5\times 10^{-17}$~erg~s$^{-1}$~cm$^{-2}$. These are galaxies a factor of approximately two fainter than galaxies previously studied at $z\sim 1.5$ (\citealt{sobral12b}).


The paper is organized as follows. In \S\ref{sec:data}, we present our galaxy sample and the emission-line extraction technique. Section~\ref{sec:theoretical} explains the necessary framework of dust extinction used in this analysis. The results are shown in \S\ref{sec:results}. Section~\ref{sec:discussion} provides a discussion of our results, their robustness, and a comparison with previous works. Finally, we summarize the main findings from our analysis in \S\ref{sec:summary}.

Throughout this paper, a standard cold dark matter cosmology is assumed, with matter density $\Omega_{m}=0.3$, vacuum energy density $\Omega_{\Lambda}=0.7$ and Hubble constant $H_{0}=70$~km~s$^{-1}$~Mpc$^{-1}$.





\section{Galaxy data set} \label{sec:data}
\subsection{Observations and data reduction}
The data set used in this work is part of the WFC3 Infrared Spectroscopic Parallel (WISP) survey. The WISP survey offers the opportunity to study the spectral properties of galaxies in a way that is unbiased by target selection. It also has unprecedented sensitivity to galaxies on the faint-end slope of the galaxy luminosity function at redshifts where the peak of the star formation in the Universe occurs (\eg \citealt{reddy08}).

A detailed description of the WISP program including technical details can be found in \citet{atek10,atek11}. Briefly, the WISP program is obtaining slitless and low resolution near-IR grism spectroscopy of more than 200 independent and high-latitude fields by observing in the pure parallel mode of the HST\footnote{In the pure parallel mode one or more instruments sample the HST focal plane while the prime program observes its planned target according to its desired visit schedule.}. The data are taken with the infrared channel of the WFC3 in a total of $\sim 730$ orbits using two grisms: G102 ($0.80\le \lambda \le 1.17~\mu$m, R $\sim 210$) and G141 ($1.11\le \lambda \le 1.67~\mu$m, R $\sim 130$).

The WISP data also include direct-imaging photometry with WFC3/IR and UVIS channels in the F475X, F600LP, F110W and F160W filters. IRAC~3.6~$\mu$m photometry from the Spitzer Space Telescope is available for a subset ($\sim 47$\%) of the initial galaxy sample described in \S\ref{sec:sample}.

The optical and near-IR photometry is obtained in matched isophotal apertures defined by the F110W imaging. The galaxies are first detected with SExtractor (\citealt{bertin96}) in the F110W band, computing the position in the sky and the aperture parameters (\ie semi-major axis and positional angle of the ellipse). These parameters are used for our own routine to extract the fluxes as follows. All four HST bands are extracted in the same elliptical apertures defined in the F110W images. The local background for each aperture is calculated within a square annulus using a mean with 3$\sigma$ clipping to determine the representative background per pixel. Then, the total background-subtracted flux is computed within the aperture by adding all the aperture pixels. The flux uncertainties are computed using Monte Carlo simulations. First, for each aperture pixel we randomly vary the flux according to a normal distribution with a width provided by an error map. Second, we repeat this process 1000 times, extracting the randomized flux in each realization. Finally, from the simulated flux distribution, the 1$\sigma$ uncertainty is calculated as the width of a Gaussian fit. Further details on the photometric analysis will be found in Bedregal et al. (in preparation).

\begin{figure}[!h]
\includegraphics[trim=1cm 0 1cm 0,clip=true,width=\columnwidth]{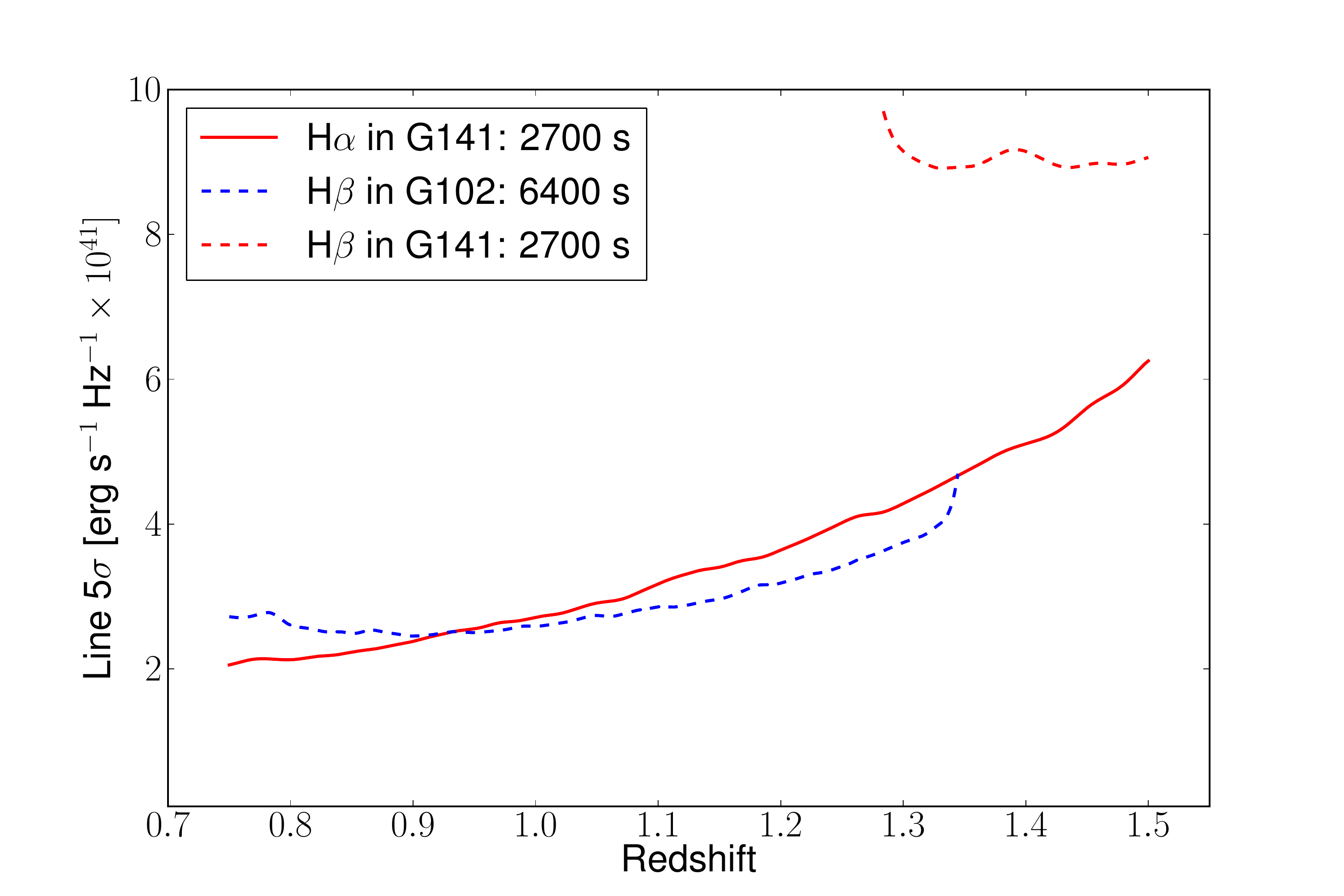}
\caption{The $5\sigma$ line sensitivity shown in \citet{atek10} is recast here as \Hb luminosity (dashed lines) and \Ha luminosity (solid line) as a function of redshift. This sensitivity corresponds to a line covering an area of 15~pixels. The blue-dashed line corresponds to the \Hb line when it is detected by the G102 grism and the red-dashed line when \Hb is detected by the G141 grism. The integration time is 6400~s for the G102 grism and 2700~s for the G141 grism. These exposure times are typical for the observations used in this study. The discontinuity in the \Hb luminosity reflects the fact that \Hb falls either in the G102 grism or in the G141 grism depending of the galaxy redshift. See \S\ref{sec:sample} for details.}
\label{fig:sensitivity}
\end{figure}

The Spitzer IRAC~3.6~$\mu$m photometry is extracted in fixed apertures of 2.8~arcsec in radius. Then, this photometry is corrected to total fluxes assuming that the galaxies are point sources\footnote{http://irsa.ipac.caltech.edu/data/SPITZER/Enhanced/Imaging/}. This is a fair assumption as the Spitzer point-spread function is significantly larger than the typical WISP aperture defined in F110W. The relatively small IRAC aperture was chosen in order to minimize any strong contamination from brighter neighbors.

The direct imaging is corrected for Galactic dust extinction given the color excess from \citet{schlegel98} and assuming the extinction curve of \citet{cardelli89}.

\subsection{Galaxy sample} \label{sec:sample}
Our initial sample contains 312 star-forming galaxies in 17 different fields in the redshift range $0.75\le z \le 1.5$, where both \Ha and \Hb fall simultaneously in the WISP spectral coverage. In the overall spectrum, our line finder flags potential emission lines as significant peaks above the continuum. The emission lines are identified and validated by eye by two astronomers, and then the redshift is assigned based on the centroid of the most significant line, which is usually H${\alpha}$ (Colbert et al., in preparation). We include in the final sample only galaxies with no external contamination (\ie galaxies with no overlapping spectra). This condition removes 49\% of the galaxies in our initial sample. This method assures that we accurately measure both continuum and emission-line fluxes that are not affected by contamination. We also make sure that only galaxies with secure redshifts are included in the final sample (that is galaxies where it is clear that the selected emission line in the spectrum is \Ha and not [\ion{O}{3}]). This contamination is avoided by looking at other emission lines at their expected wavelength. This extra condition removes another 5\% of galaxies in our initial sample. Our final sample contains 128 galaxies (with 60\% of these galaxies detected by IRAC~3.6~$\mu$m).


The WISP 5$\sigma$ line sensitivity (\citealt{atek10}) recast in luminosity is shown in Figure~\ref{fig:sensitivity} as a function of redshift. This sensitivity corresponds to a line covering an area of 15~pixels on the detector. The integration time shown in Figure~\ref{fig:sensitivity} is 6400~s for the G102 grism and 2700~s for the G141 grism. These exposure times are typical for the observations used in this study. The red-solid line is the line sensitivity for H$\alpha$, which for our redshift range is always detected in the G141 grism. However, \Hb is observed either by the G102 grism (for lower redshift galaxies) or by the G141 grism (for higher redshift galaxies). These H$\beta$-line sensitivities are shown in Figure~\ref{fig:sensitivity} with blue-dashed line for the G102 grism and with red-dashed line for the G141 grism. The \Hb line appears in both grisms at $z\sim 1.3$. At this redshift, and given the typical relative exposure times in each grism, the G141 is approximately a factor of 2.5 less sensitive than G102 in detecting H$\beta$.

\begin{figure*}
\includegraphics[width=18cm]{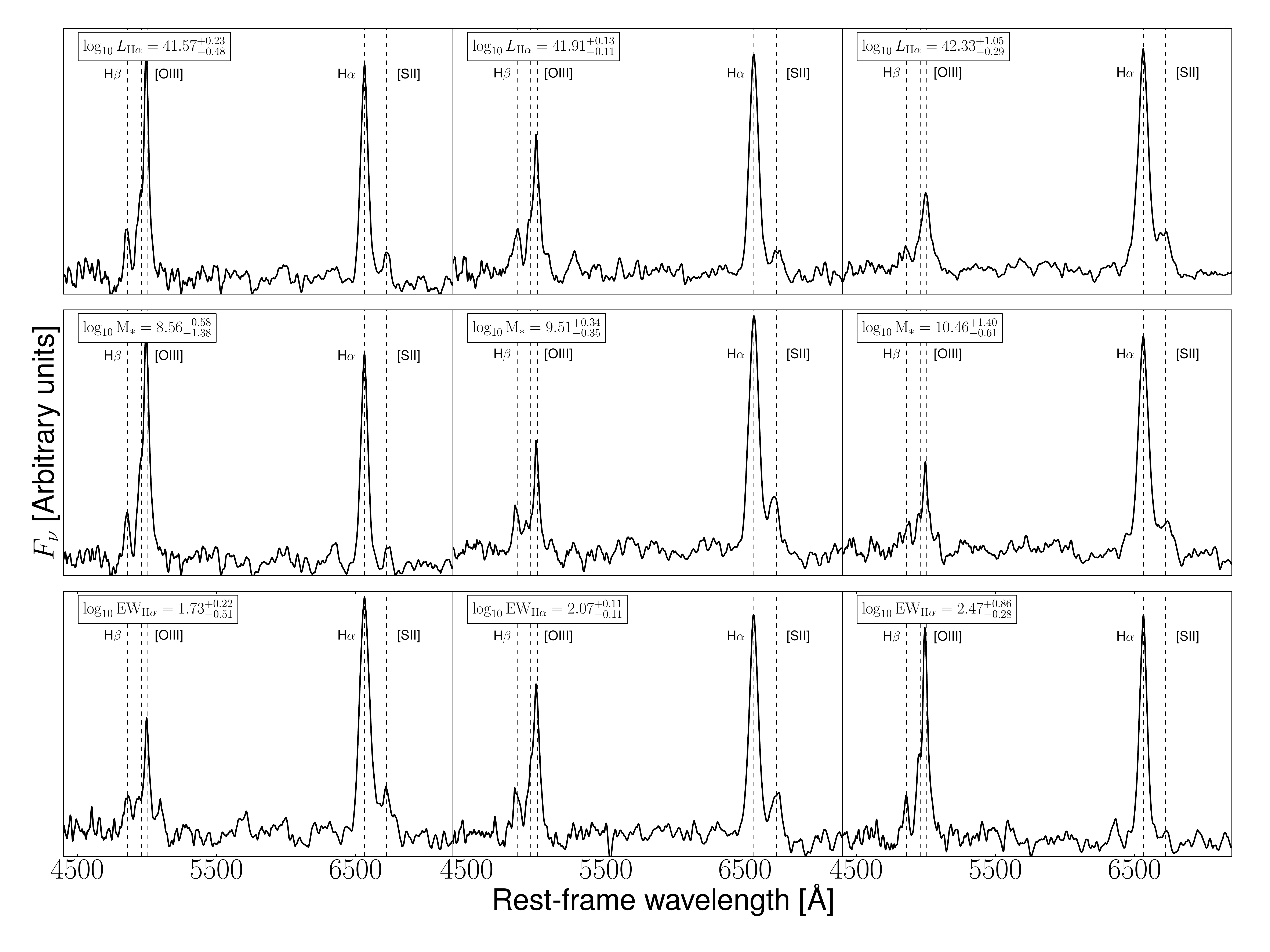}
\caption{Composite spectra in the redshift range $0.75\le z\le 1.5$. The top panels show the stacked spectra for three bins in \Ha luminosity, the middle panels show the stacked spectra for three bins in stellar mass, and the bottom a panels show the stacked spectra for three bins of rest-frame \Ha equivalent width. The number of stacked galaxies is 43, 43, and 42 in the low, middle, and high tertiles (left, center, and right columns). The emission lines that are fit in these stacked spectra are H$\beta$, the \OIII doublet, H$\alpha$ and the \SII doublet. The wavelengths of these emission lines are marked with dashed lines.}
\label{fig:compolum}
\end{figure*}

\begin{figure}[!h]
\includegraphics[trim=1cm 0 1cm 0,clip=true,width=\columnwidth]{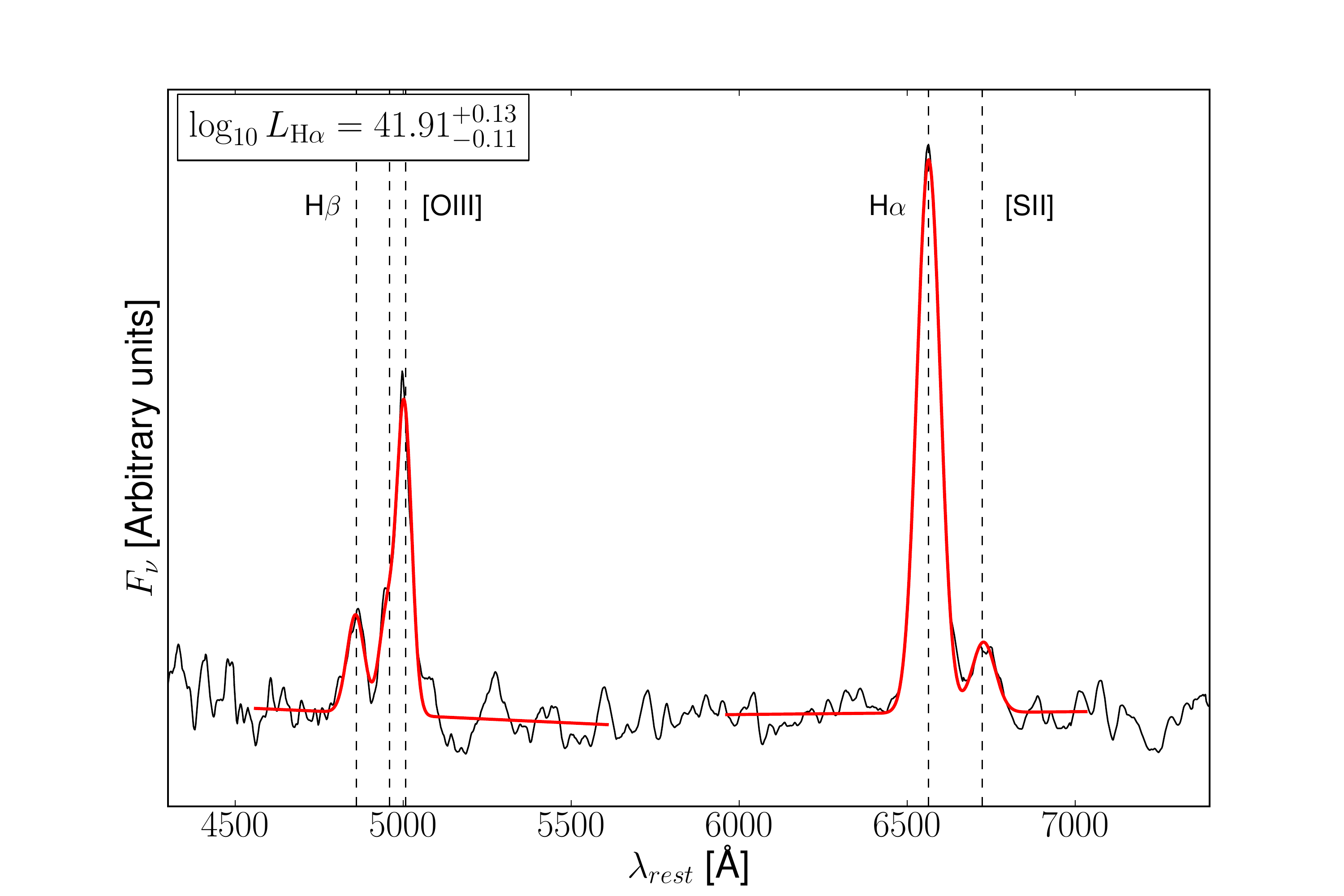}
\caption{An example of the fits to the stacked spectra. The emission-line fits are plotted with red lines. The wavelengths of the emission lines are marked with dashed lines and the observed \LHa bin is annotated in a box.}
\label{fig:example}
\end{figure}

\subsection{Emission-line extraction} \label{subsec:extraction}
A non-parametric method is developed to subtract the continuum independently from each grism. Because we know the exact wavelength of the emission line (from the automatic line detection code), we interpolate over the lines (5 pixels on either side). Then, a median filter is run with a 40 pixel window over the spectrum. This is the model of the continuum. These model continua are subtracted from the spectra.



Each spectrum is later divided into different regions in order to extract the emission-line parameters such as fluxes and equivalent widths: a bluer part containing \Hb and the \OIII doublet, and a redder part containing \Ha and the \SII doublet. In the bluer region, a function consisting of three Gaussians (one for every line in the region) plus a straight line (modeling the small residual continuum) is fit using a $\chi^{2}$ minimization. In the redder region, since our resolution is not high enough to resolve the two \SII emission lines, only two Gaussians (one for \Ha and another one for the [\ion{S}{2}] doublet) plus a straight line are fit. The \NII doublet cannot be resolved from \Ha at our resolution. The \NII contribution to the \Ha fluxes will be discussed in \S\ref{subsec:contamination}. We impose multiple constraints to the line fitting. Specifically, the full width half maximum of the Gaussians is constrained to a maximum expected size of 10 pixels ($\sim 1.3$~arcsec). The wavelength solution has an uncertainty of $\pm 30$~{\AA}. Therefore, in the fitting procedure the centers of the emission lines are allowed to vary within that wavelength range. The ratio between the fluxes of the two \OIII lines is set to 3.2 (\citealt{osterbrock89}). We note that in order to avoid obtaining local minima in our fits, we use a grid of initial conditions to find the global minimum.


Finally, Balmer decrements are computed as the ratio of the \Ha and \Hb line fluxes. The uncertainties in the Balmer decrements are then calculated by propagating the uncertainties of each of the individual line fluxes, which are obtained from the $\chi^{2}$ fits.

\subsection{Stacking of the galaxy spectra}

The individual measurements are not always good enough to detect the Balmer decrements. The \Ha emission line is detected in all galaxies by construction, while \Hb is detected at more than $3\sigma$ in 26 galaxies (this is 20\% of the final sample). Therefore, to improve the statistics, we next stack all the continuum-subtracted spectra. To stack these galaxy spectra, we first shift each spectrum to the rest-frame, interpolating the values to a finer, common wavelength grid. Different stacked spectra are built selecting the galaxies according to their \Ha luminosity, stellar mass, and \Ha equivalent width (see Figure~\ref{fig:compolum}).

The number of data points in the re-sampled spectra needs to be considered in order to correctly estimate the uncertainties in the subsequent fits. Therefore, when fitting the emission lines, the interpolated uncertainties must be corrected by a factor $\sqrt{R}$, with $R$ representing the ratio of the length of the bin in the original spectrum and the length of the bin in the interpolated spectrum. Later, every galaxy is normalized by its \Ha flux. This step is done to avoid having the stacks be dominated by the few brightest galaxies. The emission lines are closer to the edges of the stacked spectra. Thus, different lengths are fit to the left and to the right of the emission lines (300 and 600~\AA) in order to account correctly for the small residual continuum. An example of the stacked emission-line fits is shown in Figure~\ref{fig:example}. Emission-line fluxes are then extracted from these composite spectra (following the emission-line extraction procedure described in \S\ref{subsec:extraction}), which will represent statistical properties of the galaxies in every \Ha luminosity, stellar mass, and \Ha equivalent width bin. All the uncertainties are given by the python version of the \texttt{mpfit} fitting code (\citealt{markwardt09}) but we checked that those uncertainty values are compatible with uncertainties derived from a Monte Carlo simulation.

\begin{figure}[!h]
\includegraphics[trim=1cm 0 1cm 0,clip=true,width=\columnwidth]{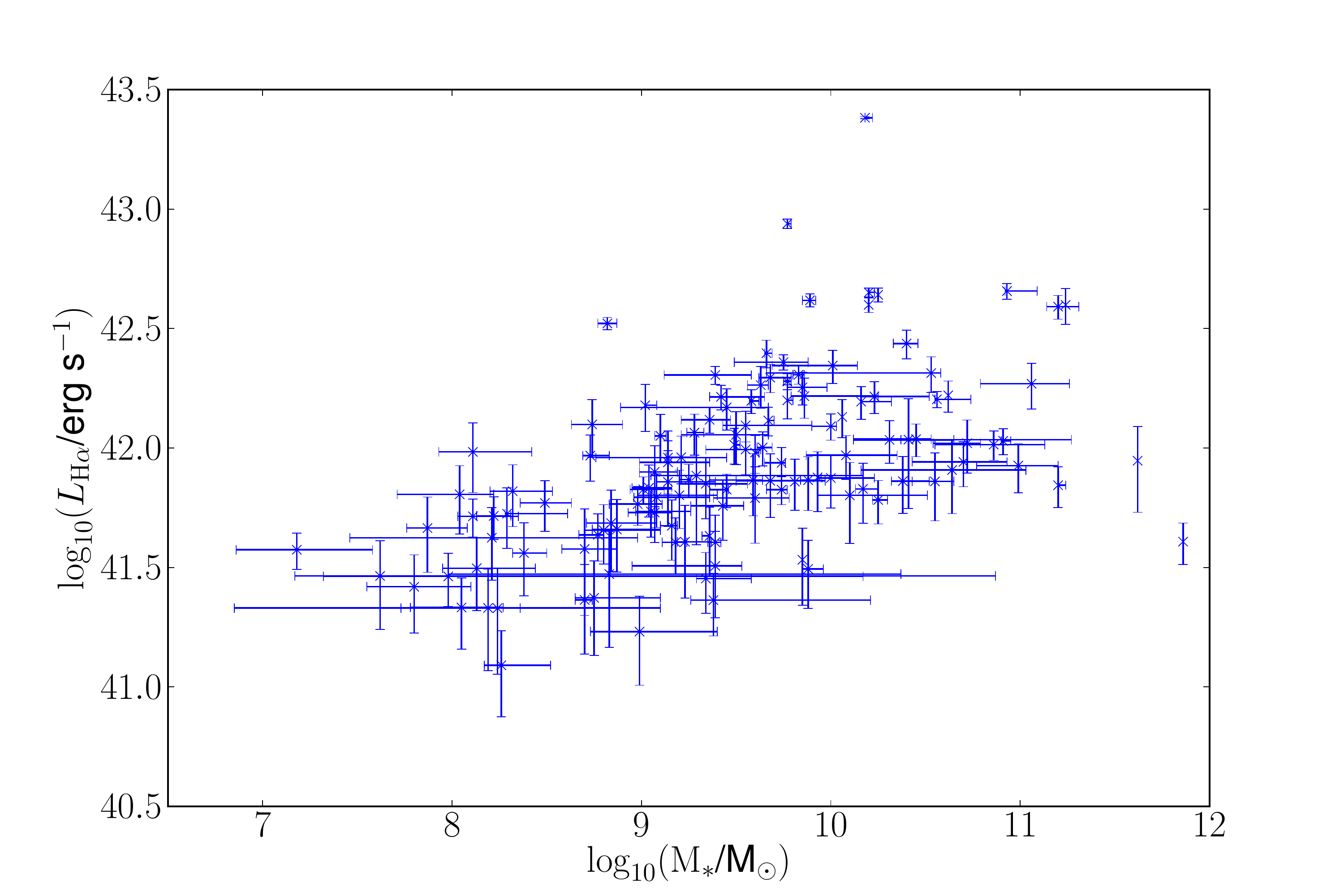}
\caption{The observed \Ha luminosity versus stellar mass of our galaxy sample.}
\label{fig:LHaobsvssmass}
\end{figure}

\begin{figure*}
\includegraphics[width=18cm]{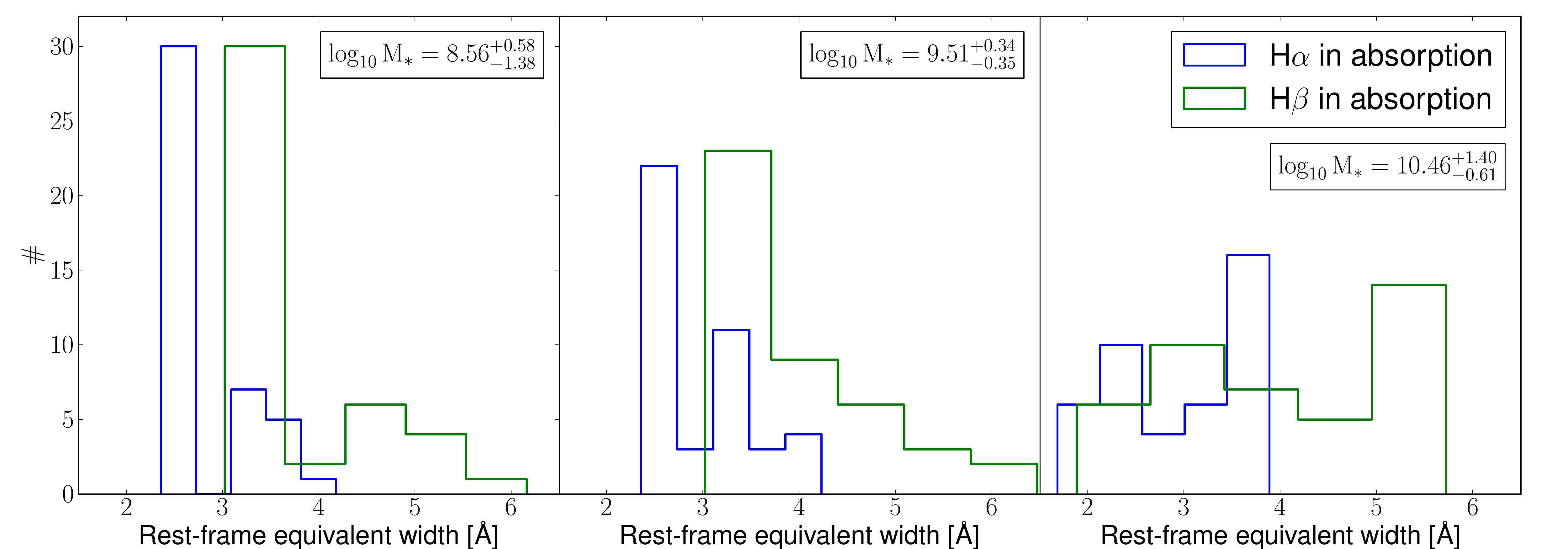}
\caption{The \Ha and \Hb absorption-line distributions of the galaxies in our sample estimated in three bins of stellar mass as described in the text.}
\label{fig:absorption}
\end{figure*}

There is the concern that the Balmer decrement determined from stacked spectra may not give the average extinction in that luminosity bin. The stacked result is dependent upon the intrinsic distribution of individual Balmer decrements, whether it is flat, gaussian or, more likely, gaussian with a tail to higher extinction. To test this potential bias, we ran a Monte Carlo simulation to make stacks from individual SDSS galaxies culled from the larger Balmer decrement distribution (in a given luminosity bin). We then measured the Balmer decrement in the stack and used the inferred extinction to determine the total intrinsic H$\alpha$ luminosity density from all of the galaxies used in the stack. The total H$\alpha$ flux was compared to the total flux when summing the H$\alpha$ fluxes corrected for extinction based on individual Balmer decrement measurements for each galaxy. The results indicate that these two methods agree within $\pm 20$\%, despite the fact that the distribution of Balmer decrements changes significantly from low to high \LHa in SDSS. These uncertainties are within the uncertainties of the Balmer decrement analyzed in the present work, so we do not consider them a significant bias.


\subsection{Galaxy stellar mass from population synthesis models} \label{subsec:SEDfitting}
Galaxy stellar mass is the most robust physical parameter derived from spectral-energy-distribution (SED) fitting. This technique has been widely used in the literature (\eg \citealt{sawicki98,papovich01,shapley01,bundy06,barro11}). In this section, we calculate stellar masses for all galaxies in our sample fitting multiwavelength photometry (F475X, F600LP, F110W, F160W and IRAC~3.6~$\mu$m; see \S\ref{sec:data}) to a grid of \citet{bruzual03} stellar population synthesis models assuming a \citet{chabrier03} initial mass function. The FAST code (\citealt{kriek09}) is used to derive these stellar masses. The FAST input parameters are: the stellar optical extinction $A^{{\rm stellar}}_{V}$ is set to the range $0-4$~mag in steps of 0.1~mag and the metallicity by mass fraction is fixed to 0.008 (solar metallicity is 0.02 in these units). The age of the galaxies is constrained in the range between $10^{7}-10^{10}$~yr and the star-formation history is modeled as an exponentially declining tau model $\exp(-t/\tau)$ with $\log_{10}(\tau/{\rm yr})=7-10$ in steps of 0.1. The contribution of the emission lines to the broad-band fluxes is removed before the SED fitting (as we suggested in \citealt{atek11}). The \Ha luminosity (uncorrected for dust absorption) versus stellar mass for galaxies in our sample is shown in Figure~\ref{fig:LHaobsvssmass}, where more luminous \Ha emission is seen in more massive galaxies. Figure~\ref{fig:LHaobsvssmass} is related to the star-formation main sequence (\eg \citealt{noeske07}; Dom\'inguez et al., in preparation).

\subsection{Contamination of Balmer line measurements} \label{subsec:contamination}
Our measurement of the Balmer emission line fluxes may be significantly affected by strong stellar absorption lines, unresolved emission lines near the Balmer lines, and contributions from AGNs.  Below we estimate the magnitude of each of these effects.

\subsubsection{Stellar absorption lines}
Absorption lines from stars in the galaxy may affect the emission-line measurements. We estimate the \Ha and \Hb absorption lines from the best stellar population model from the SED fitting described above \S\ref{subsec:SEDfitting}. The \Ha and \Hb equivalent width (EW) distributions are shown in Figure~\ref{fig:absorption} for the three galaxy mass bins that were shown in the middle panel of Figure~\ref{fig:compolum}. From the lowest to the highest mass bin, the median and standard deviation of the \Ha EWs in absorption distributions are $(2.42\pm 0.51)$~\AA, $(2.72\pm 0.52)$~\AA, $(3.05\pm 0.67)$~\AA. The corresponding values of \Hb EWs in absorption are $(3.22\pm 0.82)$~\AA, $(3.60\pm 0.87)$~\AA, $(4.09\pm 1.15)$~\AA. We compute the median of the distribution of the individual \Ha and \Hb EWs in emission that contribute to each stack. Then, the ratios between these medians and the EWs in absorption corrections are estimated. Since the fluxes are proportionally related with the EWs, we use these ratios to convert to flux corrections independently both for \Ha and H$\beta$. The uncertainties in the absorption corrections are propagated throughout our analysis.

\subsubsection{The \NII contribution to H$\alpha$}
The resolution of the WISP spectra is not high enough to resolve the \NII $\lambda$ 6548, 6583 doublet from the \Ha line. This may bias our estimate of the \Ha flux towards higher values, which may lead to systematically high Balmer decrements. Of particular concern is that the \NII contribution to \Ha is expected to increase with stellar mass (\eg \citealt{erb06}). Furthermore, the H${\alpha}$/\NII ratio at a given mass evolves with redshift (\citealt{erb06,sobral12b}).

In order to correct for this effect, we use the average of the expected H${\alpha}$/\NII observed locally (\citealt{sobral12b}) and at $z\sim 2$ (\citealt{erb06}). First, we estimate the \NII to \Ha contribution at $z\sim 2-3$ by interpolating between the H${\alpha}$/\NII ratios for different stellar masses provided in Table~2 of \citet{erb06}. These fluxes were derived for star-forming galaxies with similar physical properties as our galaxies. \citet{erb06} found that the \NII contribution relative to \Ha increases with stellar masses. For each of our stacks, the geometric mean of the stellar masses of the stacked galaxies is calculated. Then, from the previous interpolation a correction for our observed \Ha fluxes is derived. The \citet{erb06} ratios consider only the \NIIb line.  We add an extra correction to also account for the fainter (by a factor of 3) \NIIa line. The second step is to estimate the higher \NII to \Ha contribution from the relation provided by \citet{sobral12b} calibrated with local star-forming galaxies. This relation estimates H${\alpha}$/\NII as a function of \Ha EW. We proceed estimating the geometric mean of \Ha EW in our stacked spectra and apply the local correction. Finally, the total correction that we apply for the \NII contamination of \Ha is calculated as the average of these two corrections, that is from the Erb et al. (2006, approximately 0\%, 8\%, 19\% for each stellar mass bin from the least to the most massive bin) and Sobral et al. (2012, 9\%, 17\%, 20\%). The combined \NII correction has little effect on the ultimate results in this work. For instance, for the stacks binned in stellar masses these corrections are 4\%, 12\%, 19\% from the least to the most massive bin respectively.

\subsubsection{Active galactic nuclei contamination}
\begin{figure}[!h]
\includegraphics[trim=1cm 0 1cm 0,clip=true,width=\columnwidth]{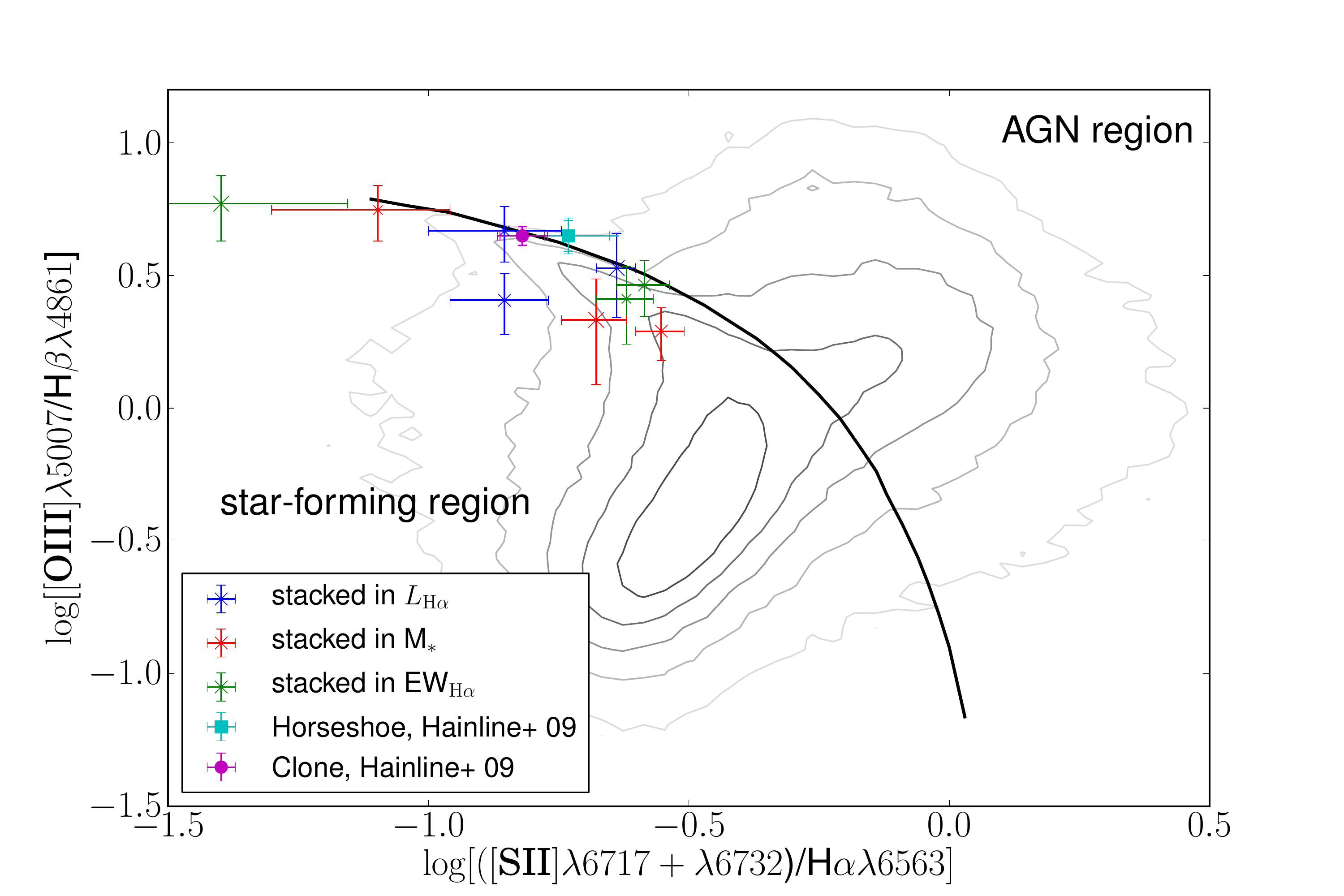}
\caption{BPT diagram that distinguishes between star-forming galaxies and galaxies with AGNs. The black line separates the star-forming and AGN regions (\citealt{kewley01}). Our 9 stacks are shown with crosses. The cross size is correlated with the bin of \Ha luminosity, stellar mass, and \Ha equivalent width (the larger the marker the larger the binning quantity). For comparison, two confirmed lensed star-forming galaxies detected at $z\sim 2$ by \citet{hainline09} are shown with a cyan square and a magenta circle. The contours are from SDSS data.}
\label{fig:bpt}
\end{figure}

We investigate whether a significant contribution from active galactic nuclei (AGN) may bias our Balmer decrement calculations. Figure~\ref{fig:bpt} shows [\ion{O}{3}]/H$\beta$ versus [\ion{S}{2}]/H$\alpha$ for our stacked spectra. This figure is the BPT diagram (\citealt{baldwin81}). The black line separates the star-forming and AGN region as modelled by \citet{kewley01}. We see in Figure~\ref{fig:bpt} that our stacks are compatible with the low-metallicity tail of the star-forming region rather than the AGN region from data in the local Universe from the SDSS DR8\footnote{http://www.sdss3.org/dr8/} (\citealt{eisenstein11}). Figure~\ref{fig:bpt} also shows for comparison the data from two lensed star-forming galaxies located at $z\sim 2$ (\citealt{hainline09}). In support of these results, the analysis by \citet{sobral12b} shows around 8\% of AGN contamination in \Ha selected samples and higher in more luminous galaxies. Since the galaxies presented in the present work are typically fainter, the AGN contamination is probably even smaller. These results indicate that AGN contamination will not significantly affect the results of our analysis. The emission-line ratios shown in Figure~\ref{fig:bpt} are listed in Table~\ref{tab1}.

\begin{deluxetable*}{c c c c}[!h]
\tablewidth{0pt}
\tabletypesize{\footnotesize}
\tablecaption{Emission-line ratios of stacked spectra in $0.75\le z \le 1.5$\label{tab1}}
\tablehead{
\colhead{$\langle \log_{10}X\rangle$} &
\colhead{([\ion{S}{2}]$\lambda 6717+\lambda 6732$)/H$\alpha \lambda 6563$} &
\colhead{[\ion{O}{3}]$\lambda 5007$/H$\beta \lambda 4861$} &
}

\startdata
\cutinhead{Stacks binned in \Ha luminosity, $X=L_{\rm H\alpha}/{\rm erg\,s^{-1}}$}
$41.57^{+0.48}_{-0.23}$ & $0.14\pm 0.04$	& $4.65\pm 1.10$\\
$41.91^{+0.11}_{-0.13}$ & $0.14\pm 0.03$	& $2.55\pm 0.66$\\
$42.33^{+0.29}_{-1.05}$ & $0.23\pm 0.02$	& $3.37\pm 1.18$\\
\cutinhead{Stacks binned in galaxy stellar mass, $X={\rm M_{*}}/{\rm M_{\odot}}$}
$8.56^{+1.38}_{-0.58}$ & $0.08\pm 0.03$	& $5.58\pm 1.32$\\
$9.51^{+0.35}_{-0.34}$ & $0.28\pm 0.03$	& $1.95\pm 0.44$\\
$10.46^{+0.61}_{-1.40}$ & $0.21\pm 0.03$	& $2.15\pm 0.92$\\
\cutinhead{Stacks binned in \Ha equivalent width, $X={\rm EW_{\rm H\alpha}}/{\rm \AA}$}
$1.73^{+0.51}_{-0.22}$ & $0.24\pm 0.03$	& $2.58\pm 0.84$\\
$2.07^{+0.11}_{-0.11}$ & $0.26\pm 0.03$	& $2.91\pm 0.69$\\
$2.47^{+0.28}_{-0.86}$ & $0.04\pm 0.03$	& $5.89\pm 1.63$
\enddata
\tablecomments{These are the values for plotting the BPT diagram shown in Figure~\ref{fig:bpt}.}

\end{deluxetable*}

\section{Theoretical background} \label{sec:theoretical}
The theoretical framework utilized in the subsequent sections is described here. Following the empirical extinction relation found in Calzetti et al. (1994\footnote{See also \citet{witt00,bell01} for their concerns about this empirical approach.}, all quantities in what follows correspond to nebular emission lines, not continuum) the intrinsic luminosities, $L_{int}$, are given by,

\[
L_{int}(\lambda)=L_{obs}(\lambda)10^{0.4 A_{\lambda}}
\]
\begin{equation} \label{eq:intrinsic0}
=L_{obs}(\lambda)10^{0.4 k(\lambda)E(B-V)}
\end{equation}
\noindent where $L_{obs}$ are the observed (extinguished) luminosities, $A_{\lambda}$ is the extinction at the wavelength $\lambda$, and $k(\lambda)$ the reddening curve. In what follows, we will use the reddening curve $k(\lambda)$ found in C00 for our analysis. The color excess $E(B-V)$ is defined by,

\begin{equation} \label{eq:ce}
E(B-V)=(B-V)_{obs}-(B-V)_{int}
\end{equation}

\noindent which is the change in the $(B-V)$ color due to dust attenuation (that is, the difference between observed and expected color in the absence of dust). The nebular color excess is related to the stellar color excess as discussed in \eg Calzetti et al. (2000, hereafter C00).


We are ultimately interested in deriving dust extinctions from Balmer decrements. The relationship between the nebular emission line color excess and the Balmer decrement is given by (see \eg the appendix in \citealt{momcheva12} for details):

\[
E(B-V)=\frac{E(H\beta-H\alpha)}{k(\lambda_{H\beta})-k(\lambda_{H\alpha})}
\]
\begin{equation} \label{eq:ce_balmer0}
=\frac{2.5}{k(\lambda_{H\beta})-k(\lambda_{H\alpha})} \log_{10} \Bigg[\frac{(H{\alpha}/H{\beta})_{obs}}{(H{\alpha}/H{\beta})_{int}}\Bigg]
\end{equation}

\noindent where $k(\lambda_{H\beta})$ and $k(\lambda_{H\alpha})$ are the reddening curves evaluated at \Hb and \Ha wavelengths, respectively. The factor $E(H\beta-H\alpha)$ is analogous to the color excess but defined for H$\beta$ and H$\alpha$, instead of the $B$ and $V$-bands. Then, $(H{\alpha}/H{\beta})_{obs}$ is the observed Balmer decrement and $(H{\alpha}/H{\beta})_{int}$ is the intrinsic or unreddened Balmer decrement, which is calculated theoretically. The intrinsic Balmer decrement remains roughly constant for typical gas conditions in star-forming galaxies (see \citealt{osterbrock89}). In our analysis, we assume the value of $(H{\alpha}/H{\beta})_{int}=2.86$, corresponding to a temperature $T=10^{4}$~K and an electron density $n_{e}=10^{2}$~cm$^{-3}$ for Case B recombination (\citealt{osterbrock89}). This choice is standard for star-forming galaxies in the literature. Thus, the nebular color excess is given by the following equation,

\begin{equation} \label{eq:ce_balmer}
E(B-V)=1.97 \log_{10} \Bigg[\frac{(H{\alpha}/H{\beta})_{obs}}{2.86}\Bigg].
\end{equation}

Coming back to Equation~\ref{eq:intrinsic0}, $A_{\lambda}$ is the extinction in magnitudes at the wavelength $\lambda$,
\begin{equation} \label{eq:intrinsic1}
A_{\lambda}=k(\lambda)E(B-V).
\end{equation}

In particular, selecting the C00 reddening curve, Equation~\ref{eq:intrinsic1} turns into,
\begin{equation} \label{eq:AHalpha}
A_{\rm H\alpha}=(3.33\pm 0.80)\times E(B-V)
\end{equation}

\begin{equation} \label{eq:AV}
A_{V}=(4.05\pm 0.80)\times E(B-V)
\end{equation}

\noindent which are the \Ha extinction and the optical extinction, respectively. The parameters $E(B-V)$, $A_{\rm H\alpha}$, and $A_{V}$ will be given in Table~\ref{tab2} for our stacked spectra. Any potential effect coming from galaxy inclinations should be alleviated from our stacking methodology since we are averaging a statistical sample of galaxies with independent inclinations.

\begin{figure*}
\includegraphics[width=18cm]{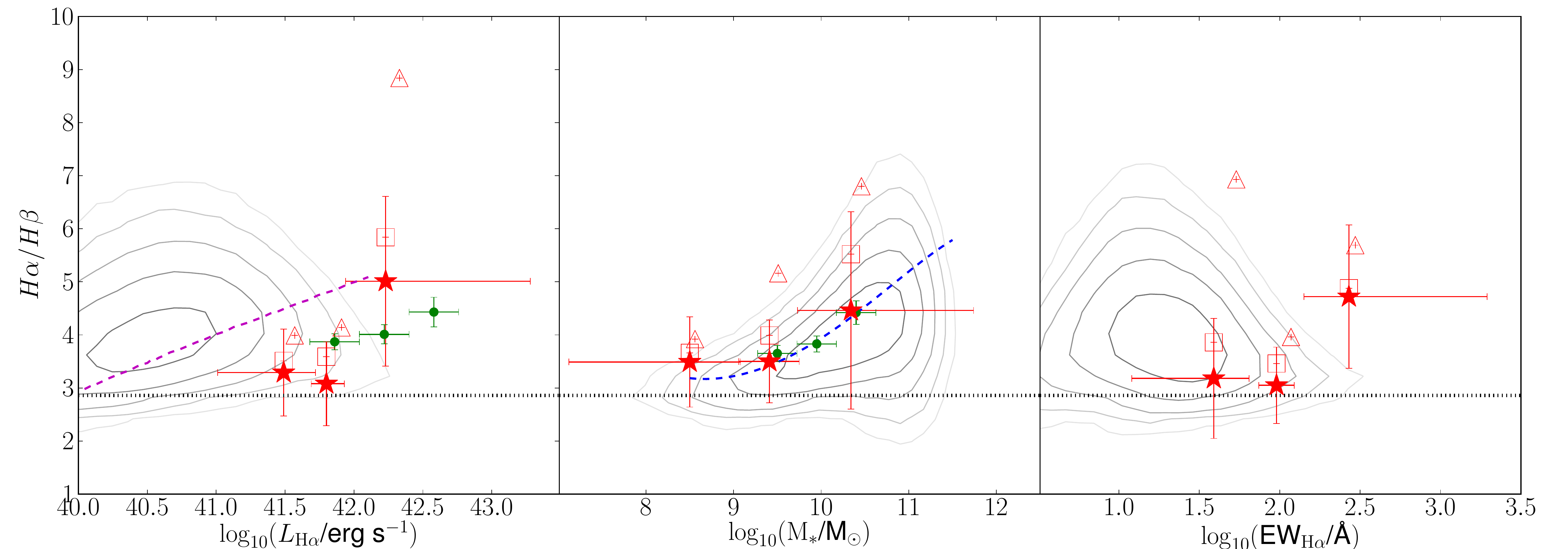}
\caption{Balmer decrements versus observed \Ha luminosity (left panel), galaxy stellar mass (middle panel), and rest-frame \Ha equivalent width (right panel). In all the panels, the open red triangles show the stacked spectra from the whole redshift range without correcting for \Ha and \Hb line absorption or \NII contamination. The open red squares are the Balmer decrements that include the absorption-line corrections (the uncertainties of the open symbols are omitted for clarity). The filled red stars show the stacked spectra from the whole redshift range $0.75\le z\le 1.5$, which are corrected for both \Ha and \Hb absorption lines and the \NII doublet contamination (see \S\ref{subsec:contamination} for details). All the values represented in this figure are listed in Table~\ref{tab2}. The dashed-magenta line in the left panel shows the results by \citet{hopkins01}. A dashed-blue line in the middle panel shows the relationship provided by \citet{garn10} for local galaxies. The results at $z\sim 1.5$ presented by \citet{sobral12b} are shown with green circles in the left and middle panels. Contours show star-forming and AGN SDSS galaxies for comparison. The dotted-black line shows the intrinsic Balmer decrement assumed in our analysis of 2.86.}
\label{fig:3in1}
\end{figure*}

\begin{figure}[!h]
\includegraphics[trim=1cm 0 1cm 0,clip=true,width=\columnwidth]{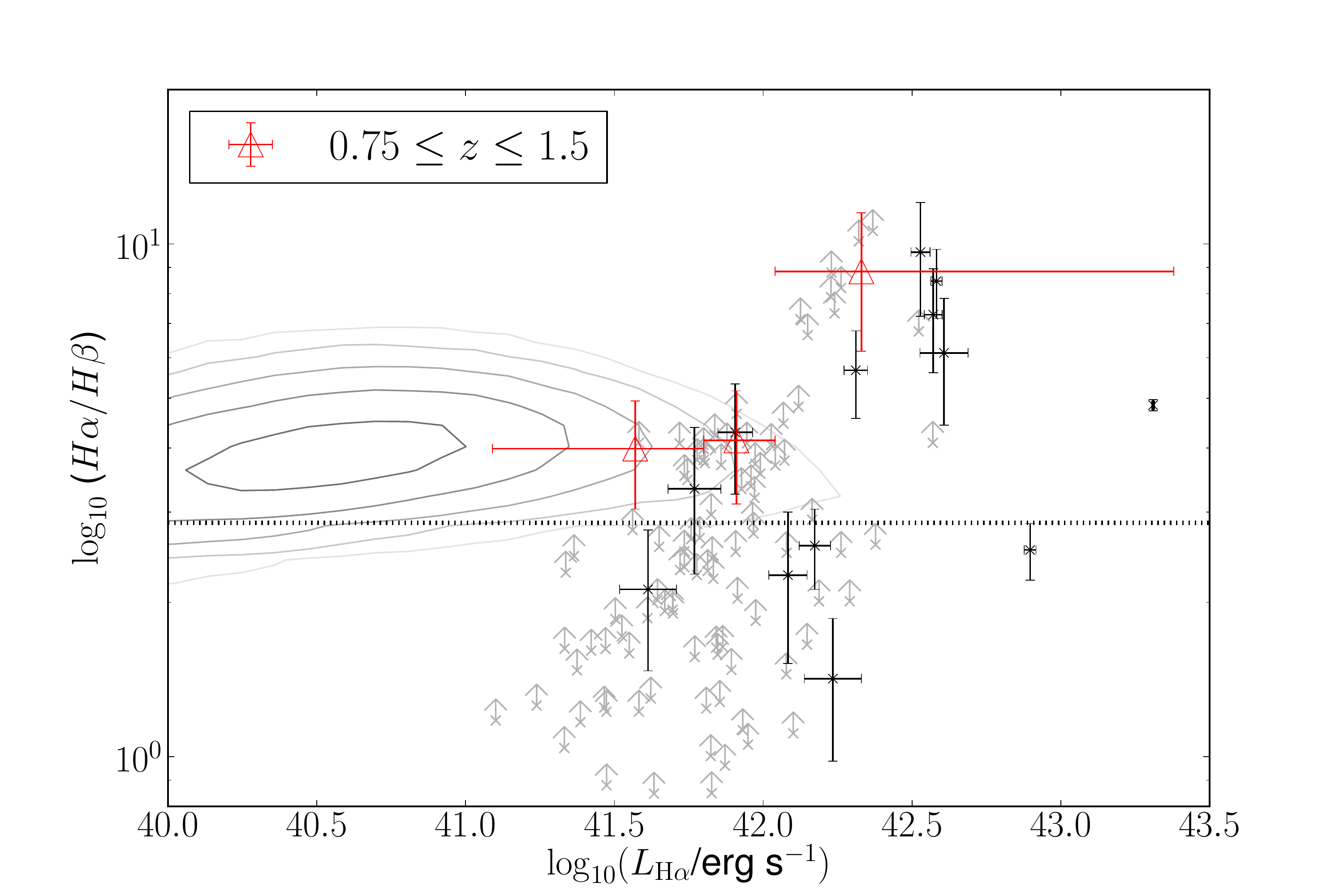}
\caption{Balmer decrement versus \Ha luminosity for individual spectra. The black crosses represent galaxies with \HaHb detected at more than 3$\sigma$ (11 galaxies). The gray arrows show 3$\sigma$ lower limits. Other symbols are the same as in the left panel of Figure~\ref{fig:3in1}.}
\label{fig:HaHbvsLHa}
\end{figure}

\section{Results} \label{sec:results}
\subsection{Balmer decrements from \Ha luminosity stacks} \label{sec:bdLHa}



Balmer decrements as a function of \Ha luminosities ($L_{{\rm H}\alpha}$) are analyzed in this section. The stacked spectra in three different observed \LHa bins are shown in the top panel of Figure~\ref{fig:compolum}. The total number of stacked galaxies is 128. The size of bins are chosen to have the same number of galaxies.

Figure~\ref{fig:3in1} (left panel) shows the Balmer decrement as a function of observed \Ha luminosity. As discussed in \S\ref{subsec:contamination}, the absorption-line corrections were applied to the \Ha and \Hb fluxes and also a correction to account for the contribution of the \NII doublet to H$\alpha$. These corrected Balmer decrements are shown with filled red stars. The open triangles show the Balmer decrements before applying any of these corrections and the open squares show the Balmer decrements after applying only the absorption-line corrections. The uncertainties are omitted in these open symbols for clarity. The intrinsic Balmer decrement considered in this work of 2.86 (dust-free galaxies) is shown with a dotted line. For comparison, star-forming galaxies in the local Universe from the SDSS are shown with contours (galaxies with AGNs were not removed from this sample). The relation between Balmer decrements and observed \LHa for these galaxies from the analysis presented in \citet{hopkins01} is shown with dashed-magenta line. The SDSS contours do not exactly match the \citet{hopkins01} results because the samples are not exactly the same. The results by \citet{ly12} at $z\sim 0.5$ are not explicitly shown in Figure~\ref{fig:3in1} (left panel) since they do not provide any parameterization. At any rate, the results by \citet{ly12} are compatible within uncertainties with the results presented in \citet{hopkins01}. The green circles in Figure~\ref{fig:3in1} are the results found in \citet{sobral12b} at $z\sim 1.5$. Table~\ref{tab2} lists the main dust properties of the stacked spectra.

\begin{figure}[!h]
\includegraphics[trim=1cm 0 1cm 0,clip=true,width=\columnwidth]{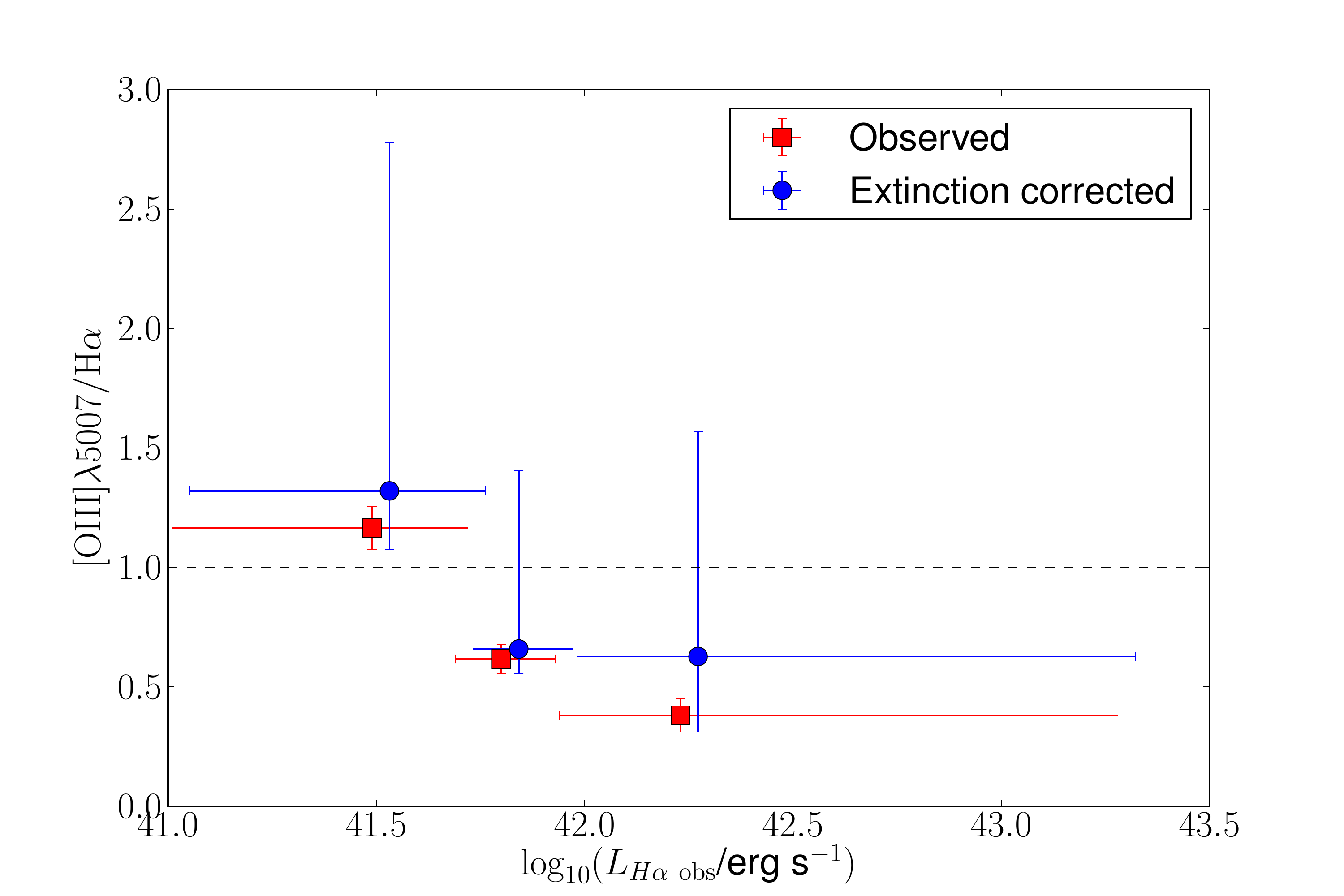}
\caption{The [\ion{O}{3}]/H$\alpha$ ratio versus \Ha luminosity. The dust-corrected data have been slightly shifted in the $x$-axis for clarity. The dashed line represents the constant value of unity.}
\label{fig:OIIIvsHa}
\end{figure}

Figure~\ref{fig:HaHbvsLHa} shows the distribution of the individual galaxies in \HaHb versus \Ha luminosity. The black crosses show galaxies with \HaHb detected at more than 3$\sigma$ whereas the gray arrows represent 3$\sigma$ lower limits. There is a broad scatter in this graph for a given observed \Ha luminosity that suggest that our extinction results can only be applied to statistical galaxy samples.

\begin{deluxetable*}{c c c c c c c c c}[!h]
\tablewidth{0pt}
\tabletypesize{\footnotesize}
\tablecaption{Dust properties for WISP galaxies\label{tab2}}
\tablehead{
\colhead{$N$\tablenotemark{a}} &
\colhead{$\langle \log_{10} (X)\rangle$} &
\colhead{$H_{\alpha}/H_{\beta}$\tablenotemark{b}} & 
\colhead{$H_{\alpha}/H_{\beta}$\tablenotemark{c}} & 
\colhead{$H_{\alpha}/H_{\beta}$\tablenotemark{d}} & 
\colhead{$E(B-V)$\tablenotemark{e}~[mag]} & 
\colhead{$A_{\rm H\alpha}$\tablenotemark{e}~[mag]} &
\colhead{$A_{V}$\tablenotemark{e}~[mag]} &
}
\startdata
\cutinhead{Galaxy spectra stacked in \Ha luminosity, $X\equiv L_{\rm H\alpha}/{\rm erg\,s^{-1}}$}
 43 & $41.57^{+0.23}_{-0.48}$ & $3.99\pm0.95$ & $3.51\pm0.85$ & $3.29\pm0.82$ & $0.12\pm0.21$ & $0.40\pm0.71$ & $0.48\pm0.86$\\
 43 & $41.91^{+0.13}_{-0.11}$ & $4.14\pm1.03$ & $3.59\pm0.91$ & $3.08\pm0.79$ & $0.06\pm0.22$ & $0.21\pm0.73$ & $0.25\pm0.89$\\
 42 & $42.33^{+1.05}_{-0.29}$ & $8.84\pm2.66$ & $5.84\pm1.86$ & $5.01\pm1.60$ & $0.48\pm0.27$ & $1.59\pm0.98$ & $1.94\pm1.17$\\
\cutinhead{Galaxy spectra stacked in stellar mass, $X\equiv {\rm M_{*}}/{\rm M_{\odot}}$}
 43 & $8.56^{+0.58}_{-1.38}$ & $3.92\pm0.93$ & $3.66\pm0.87$ & $3.49\pm0.85$ & $0.17\pm0.21$ & $0.57\pm0.71$ & $0.69\pm0.85$\\
 43 & $9.51^{+0.34}_{-0.35}$ & $5.16\pm1.10$ & $3.99\pm0.88$ & $3.50\pm0.78$ & $0.17\pm0.19$ & $0.57\pm0.65$ & $0.70\pm0.78$\\
 42 & $10.46^{+1.40}_{-0.61}$ & $6.80\pm2.79$ & $5.52\pm2.29$ & $4.46\pm1.86$ & $0.38\pm0.36$ & $1.26\pm1.22$ & $1.54\pm1.47$\\
\cutinhead{Galaxy spectra stacked in rest-frame \Ha equivalent width, $X\equiv {\rm EW_{\rm H\alpha}}/{\rm \AA}$}
 43 & $1.73^{+0.22}_{-0.51}$ & $6.93\pm2.16$ & $3.86\pm1.36$ & $3.18\pm1.13$ & $0.09\pm0.30$ & $0.30\pm1.01$ & $0.36\pm1.23$\\
 43 & $2.07^{+0.11}_{-0.11}$ & $3.96\pm0.90$ & $3.46\pm0.80$ & $3.05\pm0.72$ & $0.06\pm0.20$ & $0.19\pm0.67$ & $0.23\pm0.81$\\
 42 & $2.47^{+0.86}_{-0.28}$ & $5.69\pm1.58$ & $4.88\pm1.37$ & $4.72\pm1.35$ & $0.43\pm0.24$ & $1.42\pm0.88$ & $1.73\pm1.05$

\enddata
\tablecomments{All observables are given for nebular properties.}
\tablenotetext{a}{$N$ is the number of stacked galaxy spectra in the bin.}
\tablenotetext{b}{No correction applied.}
\tablenotetext{c}{Corrected only for \Ha and \Hb absorption lines (see \S\ref{subsec:contamination}).}
\tablenotetext{d}{Corrected for \Ha and \Hb absorption lines and \NII contamination (see \S\ref{subsec:contamination}).}
\tablenotetext{e}{Calculated from the absorption line and \NII corrected Balmer decrements.}

\end{deluxetable*}

For the faintest star-forming galaxies \OIIIb is brighter than H$\alpha$. This fact is seen clearly in Figure~\ref{fig:OIIIvsHa}, where we show the [\ion{O}{3}]$\lambda 5007$/H$\alpha$ ratio as a function of \Ha luminosity. The red squares show the observed [\ion{O}{3}]$\lambda 5007$/H$\alpha$. The blue circles show the [\ion{O}{3}]$\lambda 5007$/H$\alpha$ after correcting for dust extinction using the Balmer decrements at every observed \Ha luminosity bin. The trend of increasing [\ion{O}{3}]$\lambda 5007$/H$\alpha$ ratio can not be explained by dust extinction since the relative corrections between the two wavelengths are generally small. After correcting for dust extinction, the faintest luminosity bin has higher [\ion{O}{3}]$\lambda 5007$/H$\alpha>1$, significantly higher than in the higher luminosity bins. The relative increase in \OIIIb flux is likely due to decreasing metallicity and/or ionization parameter (\eg \citealt{kewley02}). 

\subsection{Balmer decrements from galaxy stellar mass stacks} \label{sec:bdmass}

We also analyze the Balmer decrement from the stacked spectra versus galaxy stellar mass. Figure~\ref{fig:3in1} (middle panel) shows the Balmer decrements calculated for the full galaxy sample. (The three composite spectra are shown in the middle panel of Figure~\ref{fig:compolum} as well.) The symbols are the same as in the left panel of Figure~\ref{fig:3in1}. We also show star-forming galaxies in the local Universe from the SDSS with contours (as in the previous section, galaxies with AGNs were not removed). The dashed-blue line shows the relationship between the Balmer decrement and galaxy stellar mass calculated from the analysis of a different SDSS sample by \citet{garn10}. The green circles are the results found in \citet{sobral12b} at $z\sim 1.5$. Table~\ref{tab2} lists the dust properties of the galaxy spectra stacked in stellar masses.

\subsection{Balmer decrements from \Ha equivalent width stacks}
The equivalent width of the \Ha emission line (EW$_{\rm H\alpha}$) is a proxy for the specific SFR of a galaxy. This quantity is defined as the width of the integrated continuum underneath the line that matches the flux of the line. The rest-frame \Ha EW are used in this section, this is the observed \Ha EW divided by $(1+z)$. We study as well the Balmer decrement as a function of EW$_{\rm H\alpha}$. The stacked spectra for the overall sample, binned in EW$_{\rm H\alpha}$, are shown in the bottom panel of Figure~\ref{fig:compolum}. The right panel of Figure~\ref{fig:3in1} shows the Balmer decrement calculated from the WISP stacked spectra as a function of \Ha EW. All colors and symbols are the same as in the other panels of Figure~\ref{fig:3in1}. The main properties of the \Ha EW based stacks are listed in Table~\ref{tab2}.

\section{Discussion} \label{sec:discussion}

\subsection{Balmer decrements at $0.75\le z \le 1.5$}
We find from Figure~\ref{fig:3in1} (left panel) that the Balmer decrement, and therefore dust extinction, increases with observed \LHa in the luminosity range of our analysis. The increase of the Balmer decrement with observed \LHa was already established at lower redshifts by \citet{brinchmann04,garn10,ly12}. This trend was also found recently by \citet{sobral12b} at $z\sim 1.5$ using less robust dust indicators than the Balmer decrement. \citet{sobral12b} used the ratio H$\alpha$/[\ion{O}{2}] (taken with narrow-band photometry), which is dependent on the unknown galaxy metallicity. We also note that the high star-forming galaxies represented in our sample are found rarely in the local Universe. On the other hand, galaxies with the lowest observed \LHa ($10^{41}{\rm~erg~s}^{-1}\lesssim L_{\rm H\alpha} \lesssim 10^{42}$~erg~s$^{-1}$) were found in \S\ref{sec:results} to have Balmer decrements compatible with dust-free galaxies.

Our data are not enough to characterize the dependence of the Balmer decrement with galaxy stellar mass but it is consistent with no evolution of the local relation (Figure~\ref{fig:3in1}, middle panel). A trend of the Balmer decrement with EW$_{\rm H\alpha}$ (which traces \emph{specific} star formation) is also not evident from the right panel of Figure~\ref{fig:3in1}.

\subsection{Evolution of the Balmer decrement with redshift}
There is a redshift evolution in dust properties as a function of observed \LHa from the local Universe to higher redshifts. This result remains valid even without applying the \NII corrections. Our results are compatible with the claims by \citet{sobral12b} that an $L_{*}$ galaxy\footnote{According to a \citet{schechter76} parameterization.} at $z\sim 1$ (between $10^{42.25}$ and $10^{42.50}$~erg~s$^{-1}$, \citealt{sobral12b}) has the same dust extinction as an $L_{*}$ galaxy today (approximately $10^{41.20}$~erg~s$^{-1}$, \citealt{ly07}). This result comes from the fact that the evolution in the dust-extinction dependence with \LHa is similar to the evolution from the local Universe to $z\sim 1$ of $L_{*}$ in the luminosity function. On the other hand, looking at the middle panel of Figure~\ref{fig:3in1}, we see that our galaxies span the same stellar mass range as the galaxies studied in the local Universe by the SDSS. The dependence of the Balmer decrement with stellar masses does not seem to evolve with time, which is in agreement with results by \citet{sobral12b} utilizing photometry and a different tracer of extinction.

These results are consistent with the following picture of galaxy evolution. Young star-forming galaxies have typically low content of dust and metals. Significantly older galaxies that built many stars in the past will be dustier and metal rich. When observing younger galaxies at earlier cosmic times, we see less dust in the interstellar medium even at fixed SFR (see \eg \citealt{reddy06,reddy10}). Nevertheless, the galaxy stellar mass is correlated with the production of dust and metals in such a way that the same amount of dust is produced per unit of formed stellar mass regardless of which redshift it was formed. Typically the dust is mainly produced by asymptotic giant branch (AGB) stars, which are formed when the galaxy is several 100~Myr old. Therefore, we expect that the correlation of dust extinction with stellar mass will be similar at any epoch except at higher redshifts, when most stars have not had time to enter the AGB phase. Not surprisingly, we find that this issue is not significant at the epoch around $z\sim 1$, when the universe was nearly half its present age. However, there is one caveat (which is still controversial) that we do not consider in this picture: dust may be destroyed and/or expelled from the galaxy by outflows.

A possible redshift evolution in the Balmer decrements within our sample is investigated by means of binning in two different redshifts ranges but no conclusion can be drawn from our sample due to large uncertainties.

We note that estimating dust extinction from the Balmer decrement is based on the assumption that the \Ha extinction is constant across all \Ha emitting region. As with any measure of extinction derived from unextinguished light, the Balmer decrement can not account for optically thick regions of the galaxies.  This may result in an underestimate of the total extinction (\eg \citealt{meurer02}). However, because the extinction is higher in the ultraviolet and because the intrinsic ratios are known, we believe that our Balmer line-derived estimates of the extinction are more robust than those derived with ultraviolet colors. Also, when comparing to other epochs to determine an evolution in extinction, we use the same extinction metrics for consistency (the Balmer decrement). Ideally, we would like to also measure the infrared luminosities of these galaxies to fully account for obscured star formation, but that is not possible given the resolution and confusion limits of existing infrared telescopes.

\section{Summary} \label{sec:summary}
Statistical interstellar-dust properties of star-forming galaxies were analyzed in this work as a function of $L_{\rm H\alpha}$, galaxy stellar mass, and rest-frame \Ha equivalent width. These properties were derived by stacking the spectroscopic observations of Balmer decrements of 128 star-forming galaxies from the WISP survey.

Evolution in dust-extinction properties from the local Universe up to $z=1.5$ as a function of \LHa is found as already suggested by other more model-dependent (\eg \citealt{buat07,burgarella07}) or less robust dust-extinction indicators (\eg \citealt{sobral12b}). Our data suggest that galaxies of the same luminosity are more attenuated in the local Universe than at $z\sim 1$ by a factor that is dependent on luminosity. However, the Balmer decrement as a function of galaxy stellar mass does not seem to evolve with redshift (as suggested by \citealt{sobral12b}).

We show that the fainter star-forming galaxies have brighter \OIII than \Ha emission lines. Future galaxy surveys of faint star-forming galaxies may take advantage of the fact that \OIII is typically brighter than \Ha at $L_{\rm H\alpha}<5\times10^{41}$~erg~s$^{-1}$ as it will allow for detection of fainter galaxies and/or higher redshifts than H$\alpha$-targeted surveys. 

The number of sources in the WISP catalog will increase considerably before its completion and will benefit from ongoing ground-based observations. Our dust extinction results will be updated when these observations are available. The WISP team will soon present further results on the \Ha luminosity function (Colbert et al., in preparation), massive galaxies (Bedregal et al., in preparation), mass-metallicity relation (Henry et al., in preparation) and star-forming main sequence (Dom\'inguez et al., in preparation). These results will allow us to present a comprehensive picture of the Universe at $z\sim 1$ and beyond.

\acknowledgements
The authors thank Naveen Reddy for reading and improving the paper and David Sobral for fruitful discussions. This work is based on observations made with the NASA/ESA Hubble Space Telescope, which is operated by the Association of Universities for Research in Astronomy, Inc., under NASA contract NAS 5-26555. These observations are associated with programs 11696, 12283, 12568.

\bibliographystyle{apj}

\bibliography{references}

\end{document}